\documentclass{aa}  

\usepackage{amsmath}
\usepackage{amssymb}
\usepackage{graphicx}
\usepackage{color}
\usepackage{subfigure}
\usepackage{bm}
\usepackage{textcomp,gensymb}
\usepackage{bigints}
\usepackage[dvipsnames,svgnames,table]{xcolor}
\usepackage{physics}
\usepackage{chemformula}
\usepackage{adjustbox}
\usepackage{booktabs}

\usepackage{hyperref}
\hypersetup{
    colorlinks=true,
    linkcolor=Blue,
    filecolor=Blue,
    urlcolor=Blue,
    citecolor=Blue
}

\begin{document}

   \title{Binary White Dwarfs as Gravitational Wave Sources for \textit{LISA}}

   \author{Sreeta Roy\inst{1}
     \and
          Surajit Kalita\inst{1}
          \and
            Tomasz Bulik\inst{1}
           \and   
          Dorota Gondek-Rosi\'nska\inst{1}
          }

   \institute{Astronomical Observatory, University of Warsaw, Al. Ujazdowskie 4, PL-00478 Warszawa, Poland\\
              \email{sroy@astrouw.edu.pl}\\
              \email{skalita@astrouw.edu.pl}\\
              \email{tb@astrouw.edu.pl}\\
              \email{drosinska@astrouw.edu.pl}
             }

   \date{}

 
  \abstract
   {Gravitational waves (GWs) have proven to be a powerful probe of compact binary populations. In the millihertz frequency range accessible to Laser Interferometer Space Antenna (\textit{LISA}), binary white dwarfs (BWDs) are expected to constitute a dominant source, forming both individually resolvable signals and an unresolved Galactic background.}
   {In this work, we construct a Milky Way like population model and calculate the GW background from unresolved Galactic BWD in the \textit{LISA} sensitivity range, with particular emphasis on exploring and constraining uncertainties in binary stellar evolution.}
   {We employ \textsc{compas} binary population synthesis framework to generate synthetic populations of BWD in the Milky Way. Various physically motivated evolution prescriptions and initial model parameters are used to study diverse population of BWDs. From these populations, we construct the GW background and investigate the dependence of the background spectrum on the assumptions on binary analysis. We discuss the possibility of constraints on binary evolution that \textit{LISA} GW observations may yield.}
   {We find that the shape and amplitude of the background are sensitive to key binary evolution parameters like common envelope evolution and mass-transfer efficiencies. Variations in these assumptions lead to measurable differences in the predicted background spectrum.}
   {Our results demonstrate that \textit{LISA} observations of the unresolved BWD background have the potential to constrain binary evolution models. This highlights the importance of GW background modelling as a complementary tool for studying the formation and evolution of compact binaries in the Milky Way.}


   \maketitle
%

\section{Introduction}{\label{Sec1}}

The Milky Way (MW) contains a large and diverse population of compact object binaries including white dwarfs (WDs), neutron stars (NSs), and black holes (BHs) (see \citealt{2016PhRvL.116f1102A,2023LRR....26....2A} for comprehensive reviews). These sources represent natural laboratories to test fundamental laws of physics \citep{Barack_2019}. Gravitational wave (GW) observations of these systems by ground-based detectors such as LIGO-Virgo-Kagra have opened a new window into the study of the most extreme astrophysical phenomena in our universe \citep{2016PhRvL.116f1102A,2017PhRvL.119p1101A,2019PhRvX...9c1040A,2023PhRvX..13d1039A,2004A&A...415..407B}. Such observations provide critical insights into a broad range of outstanding questions, from understanding how matter behaves at extremely high densities \citep{2007AdSpR..39..285G} to testing how gravity works in very strong fields, thereby studying the expansion of the Universe \citep{2019MNRAS.488L..35S}. At lower GW frequencies, however, the dominant Galactic compact binary population is expected to consist of binary white dwarfs (BWDs). Since the majority of stars are born with relatively low masses eventually becoming WDs and a large fraction of these reside in binary systems, our Galaxy is expected to contain hundreds of millions of BWDs \citep{2010ApJ...717.1006R,2001A&A...368..939N,2019MNRAS.490.5888L,2022MNRAS.511.5936K}. These binaries are persistent sources of low-frequency GWs, right in the range that will be detected by the Laser Interferometer Space Antenna (\textit{LISA}) \citep{2001A&A...365..491N,2019MNRAS.490.5888L,2022MNRAS.511.5936K,2023LRR....26....2A}. A few of the nearest or strongest systems, roughly $10^4$ in total will be bright enough for \textit{LISA} to pick out individually depending on the observed duration and detection threshold \citep{2017MNRAS.470.1894K,2019MNRAS.490.5888L}. The rest, however, will blend together into a persistent Galactic confusion background, a steady ``hum'' of unresolved signals dominating frequencies below a few millihertz \citep{1990ApJ...360...75H,1997CQGra..14.1439B,2017JPhCS.840a2024C,Nelemans_2009,2022MNRAS.511.5936K}.

This background contains valuable information about the Galaxy’s binary population. Its overall shape is set by the properties of the binaries themselves, their masses, orbital periods, spatial distributions, and evolutionary histories. Theoretically, a population of circular, GW driven binaries is expected to produce a stochastic background with \(\Omega_{\rm GW}(f) \propto f^{2/3}\), or equivalently \(h_c(f) \propto f^{-2/3}\), before deviations caused by Galactic structure, binary evolution, and source resolution become important \citep{2001astro.ph..8028P}. At frequencies of a few millihertz, the spectrum is expected to develop a turnover, or ``knee'', as \textit{LISA} transitions from a regime dominated by unresolved binaries to one in which an increasing fraction of sources can be resolved individually \citep{2006PhRvD..73l2001T,2017JPhCS.840a2024C,2021PhRvD.104d3019K}. The exact position and shape of this knee depend not only on \textit{LISA}’s observing strategy, but also on key aspects of binary evolution, such as the efficiency of common-envelope (CE) ejection, the stability of mass transfer (MT), and the effects of star formation history (SFH) and metallicity on the present day binary population \citep{2022MNRAS.511.5936K, Thiele_2023}.

This sensitivity makes the Galactic background a powerful diagnostic tool \citep{Nelemans_2009, 2025arXiv250622543S}. Different models of binary evolution leave distinct fingerprints in the background’s amplitude and spectral shape. By comparing simulated backgrounds under different evolutionary scenarios to what \textit{LISA} is expected to observe, we can narrow down the possible evolutionary pathways and improve our understanding of how compact binaries form and change over time. In this study, we focus on the confusion background from Galactic BWDs in the \textit{LISA} band. Our goal is to examine how changes in binary evolution physics influence the shape of this signal, and to explore how \textit{LISA}’s measurements could be used to constrain those evolutionary processes.

The remainder of this paper is organized as follows. Section \ref{Sec2} introduces the evolution of BWDs. Section \ref{Sec3} describes the \textsc{compas} population synthesis simulations and Section \ref{Sec4} describes the calculation of the GW background. Section \ref{Sec5} presents the power spectral density (PSD) estimation method and the statistical comparison of PSD shapes. Section \ref{Sec6} presents the results for the fiducial model as well as for models with varied CE, MT rate, metallicity ($Z$), and angular momentum (AM) loss prescriptions, followed by the statistical comparison between models. Section \ref{Sec7} discusses the results in comparison with previous work, interprets the physical origin of the model variations, outlines the astrophysical implications for \textit{LISA}, and summarizes the main limitations and caveats. 


\section{Binary White Dwarf evolution}{\label{Sec2}}
We model the evolution of binary systems from the zero-age main sequence (ZAMS) until both stars become WDs. The final BWD population depends on the initial stellar masses, orbital separation, eccentricity, metallicity, and on the binary interaction processes that occur during the evolution. These processes include Roche-lobe overflow (RLOF), MT, CE evolution, and AM loss. In very wide binaries, the two stars may evolve almost independently and can still form wide BWDs. However, the formation of close BWDs generally requires one or more phases of binary interaction, which can strongly change the orbital separation, component masses, and final merger time. 

\subsection{Mass Transfer by RLOF}
MT begins when one of the stars evolves and expands until its radius reaches the Roche-lobe radius. The Roche lobe is the critical equipotential region around a star in the rotating frame of the binary. Its size mainly depends on the binary separation and the mass ratio between the two stars. For a binary with semi-major axis \(a\), the Roche-lobe radius ($R_L$) of the donor can be approximated by the Eggleton relation \citep{1983ApJ...268..368E} as
\begin{equation}
\frac{R_L}{a} =
\frac{0.49 q^{2/3}}
{0.6 q^{2/3} + \ln\left(1+q^{1/3}\right)} ,
\end{equation}
where \(q=M_{\rm donor}/M_{\rm accretor}\).

Once the donor fills its Roche lobe, material can flow through the inner Lagrangian point toward the companion. The stability of this MT depends on how the donor radius and Roche-lobe radius respond during mass loss. This response is affected by the internal structure of the donor, its evolutionary stage, the mass ratio, the efficiency of accretion, and the AM carried away by any material lost from the system. If the transfer remains stable, it can proceed on nuclear or thermal timescales and may allow the accretor to gain mass while the orbit gradually changes. If the transfer becomes dynamically unstable, the MT rate increases rapidly leading to a CE phase.

\subsection{The Common Envelope Phase}
The CE phase \citep{1976IAUS...73...75P} is one of the most important but less understood stages in close binary evolution \citep{2005MNRAS.356..753N}. During this phase, the core of the donor and the companion orbit inside a shared envelope. Drag forces inside the envelope remove orbital energy and AM from the binary, causing the two cores to spiral inward. If enough energy is transferred to the envelope, it can be ejected and the system survives as a much tighter binary. If the envelope is not successfully ejected, the two cores merge and the system does not form a detached BWD.

A widely used description of CE evolution is the energy formalism, given by \cite{1984ApJ...277..355W} as
\begin{equation}
\alpha_{\rm CE}
\left(
\frac{G M_{\rm c,1} M_2}{2a_{\rm f}}
-
\frac{G M_1 M_2}{2a_{\rm i}}
\right)
=
\frac{G M_1 M_{\rm env,1}}{\lambda R_1},
\end{equation}
where $G$ is the Newton gravitational constant, \(M_1\) is the donor mass at the onset of CE, \(M_{\rm c,1}\) and \(M_{\rm env,1}\) are its core and envelope masses respectively, \(M_2\) is the companion mass, \(R_1\) is the donor radius, and \(a_{\rm i}\) and \(a_{\rm f}\) are respectively the initial and final separations. The parameter \(\alpha_{\rm CE}\) gives the efficiency with which the released orbital energy is used to eject the envelope, while \(\lambda\)  parameterizes the donor-envelope binding energy and depends on the donor’s internal structure. Because both parameters are uncertain, CE evolution remains one of the largest sources of uncertainty in binary population synthesis. In addition to the standard energy-based treatment, alternative prescriptions have been proposed. For example, the AM-based \(\gamma\)-formalism introduced by \citet{2000A&A...360.1011N} relates the change in orbital separation to AM lost from the system, rather than directly to the orbital energy budget. Therefore, different CE prescriptions can lead to different post CE orbital separations and merger time distributions for BWDs.

\subsection{Formation Pathways of BWDs}
BWDs can form through several evolutionary channels. In wide systems, both stars may evolve into WDs without strong interaction, producing wide BWDs. However, close BWDs usually require orbital shrinkage through CE evolution, stable MT, or a combination of both.

One common pathway begins with stable RLOF from the initially more massive star. The donor loses its envelope and becomes the first WD, while the companion remains non-degenerate and may gain mass. Later, the companion evolves, fills its Roche lobe, and may undergo unstable MT. This can lead to a CE phase, and if the envelope is ejected successfully, the system becomes a close BWD.

Another important pathway involves two CE phases. In this case, both MT episodes are unstable. The first CE phase produces a tighter binary containing a WD and a non-degenerate companion. The second CE phase forms the second WD and shrinks the orbit further. This channel can produce compact BWDs with short orbital periods and relatively short GW merger times. The relative importance of these formation channels depends strongly on the adopted assumptions for MT stability, CE efficiency, accretion efficiency, and AM loss.

\subsection{Gravitational Wave Driven Evolution of BWDs}
Once a close BWD system has formed, its subsequent evolution is dominated by the emission of GWs. In general relativity, an accelerating binary loses energy and AM through GWs. For a detached binary in a nearly circular orbit, this loss causes the orbital separation to decrease over time, shortening the orbital period and shifting the emitted GWs to higher frequencies. For a binary in a circular orbit, the orbital decay rate due to GW emission is given by \cite{1964PhDT........51P} as
\begin{equation}\label{Eq: Orbital decay}
\dv{a}{t} = -\frac{64}{5} \frac{G^3 M_1 M_2 (M_1+M_2)}{c^5 a^3},
\end{equation}
where $c$ is the speed of light. This expression shows that the inspiral accelerates as the binary shrinks.

Using Kepler’s law, the orbital evolution can equivalently be expressed in terms of the orbital period ($P$) as
\begin{equation}
\dv{P}{t} = -\frac{96}{5} \frac{G^{5/3}}{c^5} \frac{(2\pi)^{8/3} M_1 M_2}{(M_1+M_2)^{1/3}} P^{-5/3}.
\end{equation}
For circular binaries, the GW frequency is twice the orbital frequency; thus a decreasing $P$ corresponds to an increasing GW frequency. Since many BWDs evolve slowly in the mHz band, they can remain nearly monochromatic over the \textit{LISA} mission lifetime, although the shortest-period and most massive systems may exhibit significant frequency evolution.

At leading quadrupole order, the GW strain amplitude ($h_0$) from a circular detached BWD depends on the chirp mass ($\mathcal{M}_c$), GW frequency ($f$), and distance to the source ($D$) as \citep{2012A&A...544A.153S}
\begin{equation}
h_0(f) = \frac{4 (G\mathcal{M}_c)^{5/3} (\pi f)^{2/3}}{c^4 D},
\end{equation}
with
\begin{equation}
    \mathcal{M}_c = \frac{(M_1 M_2)^{3/5}}{(M_1+M_2)^{1/5}},
\end{equation}
The prefactor 4 corresponds to the standard quadrupole-order expression for the intrinsic strain amplitude of a circular binary. Different numerical prefactors may appear in the literature depending on whether one quotes the instantaneous or time-averaged strain, the individual polarization amplitudes or the combined signal, and whether the result is given for an optimally oriented source or averaged over inclination, sky position, polarization, and detector response.

As GWs carry away orbital energy and AM, the two WDs spiral closer together, causing the orbital period to shorten and the emitted GW frequency to increase by chirping. The rate of this chirp depends strongly on the system’s masses and separation. Ultra compact binaries with periods of just a few minutes evolve rapidly, and \textit{LISA} will be able to detect measurable frequency changes over its mission lifetime. Such measurements provide direct constraints on $\mathcal{M}_c$ and $\dot{a}$, offering a test of GW emission from these binaries. In contrast, wider BWD systems evolve much more slowly. Over the timescale of \textit{LISA} observations, their frequencies remain effectively constant, and they contribute long lived, nearly monochromatic signals. These slowly evolving binaries are the dominant contributors to the unresolved Galactic background, forming the quasi steady background ``hum'' in the millihertz band. The proportion of rapidly evolving to slowly evolving systems determines not only the slope of the low frequency tail but also how sharply the background drops above the turnover frequency, where the number of resolvable sources begins to outnumber the unresolved population.

The present day distribution of BWD orbital periods is likely a direct product of GW driven orbital decay acting on the population produced by earlier binary evolution processes such as MT and CE ejection. This makes GW driven evolution a critical link between binary evolution physics and the observable properties of the \textit{LISA} background. Variations in the efficiency of CE ejection, the stability of MT, and the star formation history influence how quickly binaries evolve under GW emission; thus leaving measurable imprints on the background signal's amplitude, spectral slope, and knee frequency.

In this work, we focus on studying these imprints. By modelling how GW driven evolution shapes the unresolved signal in the \textit{LISA} band and by comparing different binary evolution scenarios, we aim to determine which features of the observed background can be used to study the physics of BWD formation and evolution.


\section{Utilizing \textsc{COMPAS} simulation}{\label{Sec3}}
To generate our synthetic population of BWDs, we use \textsc{compas} v02.50.00 \citep{2022ApJS..258...34R}, an open-source rapid binary population synthesis (BPS) code. \textsc{compas} follows the evolution of binary systems using parametrised prescriptions for stellar evolution, MT, CE evolution, compact-object formation, and AM loss. Similar to other rapid BPS codes such as \textsc{bse}~\citep{2002MNRAS.329..897H}, \textsc{StarTrack}~\citep{2008ApJS..174..223B}, \textsc{Binary C}~\citep{2004MNRAS.350..407I}, \textsc{SeBa}~\citep{2012A&A...546A..70T}, and \textsc{cosmic}~\citep{2020ApJ...898...71B}, \textsc{compas} is computationally efficient and allows the evolution of millions of binaries over a wide range of initial conditions. This makes it suitable for statistical studies of binary populations, although the detailed internal stellar structure is not resolved in the same way as in 1D stellar evolution codes such as \textsc{mesa} or other hydrodynamic simulations \citep{2018PASA...35...31P}. \textsc{compas} has mainly been used in studies of compact remnants such as neutron stars and black holes \citep{2022MNRAS.516.5737B,2022MNRAS.517.4034S,2022ApJ...937..118W}. Recent population synthesis studies have also investigated the formation channels and intrinsic properties of BWD populations, including their masses, separations, core compositions, and MT histories \citep{Roy_Kalita_2026}. In this study, we focus on the GW background produced by Galactic BWDs and examine how different binary evolution assumptions change its shape and amplitude in the \textit{LISA} band.

We generate a large population of binary systems by assigning initial parameters from standard astrophysical distributions. The primary masses are drawn from a Salpeter initial mass function in the range \(0.5-15\,{\rm M}_{\odot}\), while the secondary masses are assigned using a flat mass-ratio distribution \citep{1955ApJ...121..161S}. The semi-major axes are sampled from a distribution spanning \(0.01-10\,{\rm AU}\), and the eccentricities ($e$) are drawn from a thermal distribution, \(f(e)=2e\), in the range \(0-1\) \citep{1975MNRAS.173..729H}. For our main simulation set, we evolve \(N=10^7\) binaries using the default rapid population synthesis mode of \textsc{compas}. After the evolution, we select systems in which both components have become WDs with core comprising of carbon-oxygen (C/O), oxygen-neon (O/Ne), or helium-core (He) \citep{2002MNRAS.329..897H}. Nearly \(10^6\) systems satisfy our BWD selection criteria. We then construct different evolutionary models by varying the main uncertain binary evolution parameters. These include the CE efficiency parameter \(\alpha_{\rm CE}\), the envelope-structure parameter \(\lambda_{\rm CE}\), the MT efficiency parameter \(\beta\), metallicity $Z$, and AM loss prescriptions. For all models, we assume a uniform star-formation rate (SFR). The resulting synthetic BWD populations are then used as the input for the GW background calculations described in the following sections.


\section{Calculation of Gravitational Wave background}{\label{Sec4}}    
To compute the present day Galactic GW background from BWDs, we post-process the synthetic binary populations generated with \textsc{compas} using a custom \texttt{Python} pipeline.

We begin with the population of binaries at the epoch when they form BWD systems, extracting $a$, $e$, $M_1$, $M_2$, and formation times from the \textsc{compas} output. Galactic positions are assigned using a precomputed spatial distribution concentrated in the thin disc of Milky Way \citep{2020MNRAS.492.4043C} and heliocentric distances are obtained by placing the Sun at a Galactocentric radius of 8.5\,kpc. We assume a Galactic age of 13.6\,Gyr, with star formation occurring over the last 10\,Gyr.. To approximate continuous star formation, the Galactic history is divided into 10 equal time intervals and one burst is placed randomly within each interval.

Close BWDs emerge from mass transfer and CE evolution with negligible eccentricity due to tidal circularization during the interaction phases. We therefore set $e=0$ at BWD formation and evolve each orbit from its formation time to the present epoch using Equation~\eqref{Eq: Orbital decay}. Systems that merge before the present day are removed from the sample. For the surviving binaries, we compute the present day orbital separation, thereby evaluating their GW properties in the \textit{LISA} band. From the evolved $a$, $M_1$, and $M_2$, we compute the current orbital frequency using Kepler’s law. For each source, we now define
\begin{equation}
   A = \frac{(G\mathcal{M}_c)^{5/3} (\pi f)^{2/3}}{c^4 D}.
\end{equation}

For the standard quadrupole polarizations
\begin{align}
h_{+}(t)&=2A(1+\cos^{2}\iota)\cos\Phi(t),\\
h_{\times}(t)&=-4A\cos\iota\,\sin\Phi(t),
\end{align}
with $\Phi(t)$ being the phase of the signal at time $t$ and $\iota$ the inclination angle to the source, the phase-averaged squared strain is given by
\begin{equation}
\langle h^{2}\rangle_{\Phi}=\frac{1}{2}h_{+,0}^{2}+\frac{1}{2}h_{\times,0}^{2}=2A^{2}\left[(1+\cos^{2}\iota)^{2}+4\cos^{2}\iota\right].
\end{equation}
Averaging further over inclination, with \(\mu=\cos\iota\) uniformly distributed in \([-1,1]\), gives
\begin{equation}
\langle h^{2}\rangle_{\Phi,\iota}=2A^{2}\left\langle 1+6\mu^{2}+\mu^{4}\right\rangle.
\end{equation}
Using $\langle \mu^{2}\rangle=1/3$ and $\langle \mu^{4}\rangle=1/5$, it simplifies to
\begin{equation}
\langle h^{2}\rangle_{\Phi,\iota}=\frac{32}{5}A^{2}.
\end{equation}
Therefore, the root-mean-square strain becomes
\begin{equation}
h(f)=\sqrt{\langle h^{2}\rangle_{\Phi,\iota}}=\sqrt{\frac{32}{5}}\, \frac{(G\mathcal{M}_{c})^{5/3}}{c^{4}D}(\pi f)^{2/3}.
\end{equation}
We classify a source as resolved or unresolved using a monochromatic signal-to-noise ratio (SNR) estimate based on the analytic \textit{LISA} sensitivity model of  \citet{2019CQGra..36j5011R}. Assuming an observation time of \(T_{\rm obs}=4\,{\rm yr}\), we compute
\begin{equation}
{\rm SNR} \approx \frac{h(f)\sqrt{T_{\rm obs}}}{\sqrt{S_n(f)}},
\end{equation}
and adopt a threshold of \({\rm SNR}=7\). Only systems below this threshold are included in the unresolved background. We use an approximate monochromatic SNR criterion to separate resolved and unresolved systems, rather than performing a full \textit{LISA} parameter-estimation or source subtraction analysis.

To construct the effective background spectrum, we
\begin{itemize}
    \item apply the resolved cut and weigh the remaining unresolved strain power by a population scale factor so that the simulated sample represents an effective Galactic BWD population of $3\times10^8$ BWDs.
    \item bin unresolved systems in logarithmically spaced frequency bins between $10^{-4}$\,Hz and $10^{-1}$\,Hz using 15 bins per decade.
    \item sum the squared strain amplitudes of all contributing unresolved systems in each frequency bin and normalize the result by the bin width. The effective strain is then calculated as
    \begin{equation}
        h_{\mathrm{eff}}(f) = \sqrt {{f}{\frac{\sum_i h^2(f_i)}{\Delta f}}}.
    \end{equation}
\end{itemize}

The resulting $h_\mathrm{eff}(f)$ represents the effective unresolved Galactic background for a given realization and model. For each population model, the final background curve is obtained by averaging $h_\mathrm{eff}(f)$ over 100 realizations. This procedure is repeated for the fiducial population and for each set of parameter variations that include changes to CE efficiency, MT efficiency, metallicity, and AM loss prescriptions. This enables a direct comparison of how binary evolution physics shapes both the amplitude and frequency dependence of the \textit{LISA} Galactic background.


\section{Power Spectral Density Estimation}{\label{Sec5}}

To characterize the unresolved Galactic BWD signal in a form that is directly comparable to the instrumental noise budget of \textit{LISA}, we estimate its one sided strain power spectral density (PSD), \(S_h(f)\), from synthetic time-domain realizations of the source population. The present-day Galactic BWD population used for the PSD calculation is constructed in the same manner as described in Section~\ref{Sec4}. Resolved and unresolved sources are identified using the same monochromatic SNR criterion adopted for the \(h_{\rm eff}\) calculation, based on the analytic \textit{LISA} sensitivity model of \citet{2019CQGra..36j5011R}, an observing time \(T_{\rm obs}=4\,{\rm yr}\), and a threshold \({\rm SNR}<7\) for unresolved binaries. Only unresolved systems in the frequency range \(10^{-4}\,{\rm Hz}\leq f \leq 10^{-1}\,{\rm Hz}\) are retained. The unresolved source amplitudes are then rescaled so that the resulting strain power corresponds to an effective MW population of \(3\times10^8\) BWDs, consistent with the normalization adopted in Section~\ref{Sec4}.

\subsection{Time-domain representation of unresolved sources}
For the PSD construction, each unresolved source is represented as a monochromatic sinusoid with a random initial phase. In this step, we use the strain amplitude averaged over an isotropic distribution of binary orientations, but not over time, so that
\begin{equation}
h(f)=\sqrt{\frac{64}{5}}\,
\frac{(G\mathcal{M}_c)^{5/3}}{c^4D}(\pi f)^{2/3}.
\end{equation}
The prefactor \(\sqrt{64/5}\) appears because the time averaging is carried out subsequently by the PSD estimator itself. Each unresolved binary is therefore modelled as
\begin{equation}
h_i(t)=h_{i}\cos\!\left(2\pi f_i t+\phi_i\right),
\end{equation}
where \(f_i\) is the source frequency and \(\phi_i\) is a random phase drawn uniformly from \([0,2\pi]\). The total strain is the superposition of all unresolved binaries,
\begin{equation}
h(t)=\sum_i h_i(t).
\end{equation}

In practice, we construct the signal in the frequency domain on the discrete Fourier grid and then transform it into the time domain. For an observation sampled at cadence \(\Delta t\), the sampling frequency is $f_s = 1/\Delta t$ and for a segment of \(N_{\rm FFT}\) samples the Fourier bin spacing is $\Delta f = f_s/N_{\rm FFT}$. Each source frequency is assigned to the nearest discrete Fourier bin. This bin centred monochromatic approximation provides an efficient representation of the unresolved background for the purpose of estimating the ensemble PSD.

\subsection{PSD estimator}

The one sided strain PSD is estimated further. We adopt a sampling frequency of $f_s = 4\,{\rm Hz},$ an FFT length $N_{\rm FFT}=2^{17}$ and a segment overlap of 50\%. For an observing time \(T_{\rm obs}\), the total number of samples in the realization is $N_{\rm tot}=T_{\rm obs}f_s$, with hop size $N_{\rm hop}=N_{\rm FFT}(1-\zeta)$, where \(\zeta=0.5\) is the fractional overlap, and the number of segments is
\begin{equation}
K = 1 + \left\lfloor \frac{N_{\rm tot}-N_{\rm FFT}}{N_{\rm hop}} \right\rfloor .
\end{equation}
For each segment, we apply a Hann window \(w_n\) and compute the discrete Fourier transform of the windowed strain. The window normalization factor is
\begin{equation}
U = \frac{1}{N_{\rm FFT}}\sum_{n=0}^{N_{\rm FFT}-1} w_n^2.
\end{equation}
The one sided periodogram estimate for a segment is thus given by
\begin{equation}
\hat{S}_h(f_k)=\frac{2}{f_s N_{\rm FFT} U}\left|\tilde{h}_k\right|^2,
\qquad k=1,\dots,\frac{N_{\rm FFT}}{2}-1.
\end{equation}
The final PSD estimate for one realization is obtained by averaging the periodograms over all \(K\) overlapping segments
\begin{equation}
S_h(f_k)=\frac{1}{K}\sum_{j=1}^{K}\hat{S}_{h,j}(f_k).
\end{equation}
This procedure naturally performs the time averaging of the injected monochromatic signals and yields a one sided PSD in units of \({\rm strain}^2\,{\rm Hz}^{-1}\).

\subsection{Ensemble averaging and logarithmic rebinning}
For each population model, we generate 100 independent Galactic realizations and compute the PSD for each realization separately. The resulting ensemble of PSDs is then averaged to obtain the mean background PSD
\begin{equation}
\langle S_h(f_k)\rangle = \frac{1}{N_{\rm real}}\sum_{r=1}^{N_{\rm real}} S_h^{(r)}(f_k),
\end{equation}
where \(N_{\rm real}=100\). We also compute the realization to realization scatter for later statistical analysis.

To suppress small scale fluctuations and facilitate model comparison, the PSDs are logarithmically rebinned over the interval \(10^{-4}\) to \(10^{-1}\,{\rm Hz}\) using 15 bins per decade. Within each logarithmic bin, the rebinned PSD is taken as the arithmetic mean of the PSD samples falling inside that bin. The resulting rebinned spectrum provides a smooth estimate of the unresolved Galactic background in the same units $S_h(f)\quad [{\rm strain}^2\,{\rm Hz}^{-1}]$, which can be compared directly with the \textit{LISA} instrumental noise PSD.

This PSD based construction complements the effective characteristic strain approach described in the previous section. In contrast, the former provides an intuitive description of the background amplitude and the PSD estimate offers a direct spectral characterization of the unresolved Galactic signal in the form most relevant for detector noise comparisons and subsequent statistical analyses.

\subsection{Statistical comparison of PSD shapes}

To quantify differences among population models, we compare the rebinned PSDs using a $\chi^2$ statistic defined in logarithmic PSD space. This comparison is performed after computing the PSD and rebinning the resulting spectra onto a common logarithmic frequency grid.

For each model and each realization \(r\), let \(S_h^{(r)}(f_k)\) denote the rebinned PSD in frequency bin \(f_k\). Each alternative model is compared against a chosen baseline model using matched ensembles generated with the same set of random seeds, so that realization to realization fluctuations are sampled consistently across models.

Because the PSD spans several orders of magnitude, we perform the comparison in log space. For each realization \(r\) and bin \(f_k\), we define the log-difference
\begin{equation}
D_r(f_k)=\ln S_{h,{\rm model}}^{(r)}(f_k)-\ln S_{h,{\rm base}}^{(r)}(f_k).
\end{equation}
From these quantities we compute the mean log difference across the ensemble,
\begin{equation}
\bar{D}(f_k)=\frac{1}{N_{\rm real}}\sum_{r=1}^{N_{\rm real}} D_r(f_k),
\end{equation}
together with the realization to realization standard deviation
\begin{equation}
\sigma_D(f_k)=
\sqrt{
\frac{1}{N_{\rm real}-1}
\sum_{r=1}^{N_{\rm real}}
\left[D_r(f_k)-\bar{D}(f_k)\right]^2
}.
\end{equation}

The $\chi^2$ statistics is then constructed from the mean log differences over the selected frequency bins
\begin{equation}
\chi^2=\sum_{k\in\mathcal{B}} w_k\,\bar{D}^2(f_k),
\end{equation}
where \(\mathcal{B}\) is the set of frequency bins included in the comparison and \(w_k\) is the inverse variance weight in bin \(k\). We use the realization to realization scatter
\begin{equation}
w_k=\frac{1}{\sigma_D^2(f_k)},
\end{equation}
which provides a more conservative estimate of distinguishability by incorporating the intrinsic variance between different Galactic realizations.

The comparison is restricted to the subset of rebinned frequency bins in which the baseline background exceeds a chosen fraction of the \textit{LISA} instrumental noise PSD. Writing the mean baseline PSD as \(\bar{S}_{h,{\rm base}}(f_k)\) and the \textit{LISA} noise PSD as \(S_n(f_k)\), the selected bins satisfy $\bar{S}_{h,{\rm base}}(f_k)>\eta\,S_n(f_k)$, where \(\eta\) is a threshold parameter. In this work we adopt \(\eta=1\), so that the $\chi^2$ is evaluated only in the frequency interval where the mean Galactic background exceeds the instrumental noise level.

If \(N_{\rm bin}\) frequency bins satisfy this criterion, then the number of degrees of freedom is simply ${\rm dof}=N_{\rm bin}$, and we report the reduced $\chi^2$ as $\chi_\mathrm{red}^2=\chi^2/{\rm dof}$. With this weighting, the statistic measures whether the difference between two models exceeds the typical realization to realization scatter, and values well below unity are expected when the models are indistinguishable. This procedure provides a quantitative measure of how strongly the PSD predicted by a given population model differs from that of the baseline model.

\section{Results}{\label{Sec6}}

\subsection{Fiducial model}

We begin by presenting the GW background predicted by our fiducial binary population synthesis model. This model adopts the default \textsc{compas} parameters, which represent commonly used assumptions for binary stellar evolution. We assume a CE efficiency of $\alpha_\mathrm{CE} = 1.0$ with the default structural parameter $\lambda_\mathrm{CE} =0.1$. The accretion efficiency of stable MT is set by the thermal timescale limit and solar metallicity ($Z = 0.014$) is adopted. The initial parameter distributions follow those described in \textsc{compas} section and we assume a constant SFR over the age of the Galaxy. The corresponding product $\alpha_\mathrm{CE}\lambda_\mathrm{CE}=0.1$ lies within the standard range ($0.1-0.5$) explored in previous population synthesis works \citep{2001A&A...365..491N,2012A&A...546A..70T,2019MNRAS.490.5888L}.

Using these parameters, we evolve $10^7$ initial binaries, of which around  $10^6$ form BWDs by the present day. The resulting population is dominated by carbon-oxygen (C/O) and helium-core (He) white dwarfs, with a smaller fraction of oxygen-neon (O/Ne) systems. The orbital period distribution at the present epoch peaks around a few hours, with a significant tail extending down to more compact systems that emit in the high-frequency end of the \textit{LISA} band.

\begin{figure*}[htpb]
    \centering
    \includegraphics[width=0.78\textwidth]{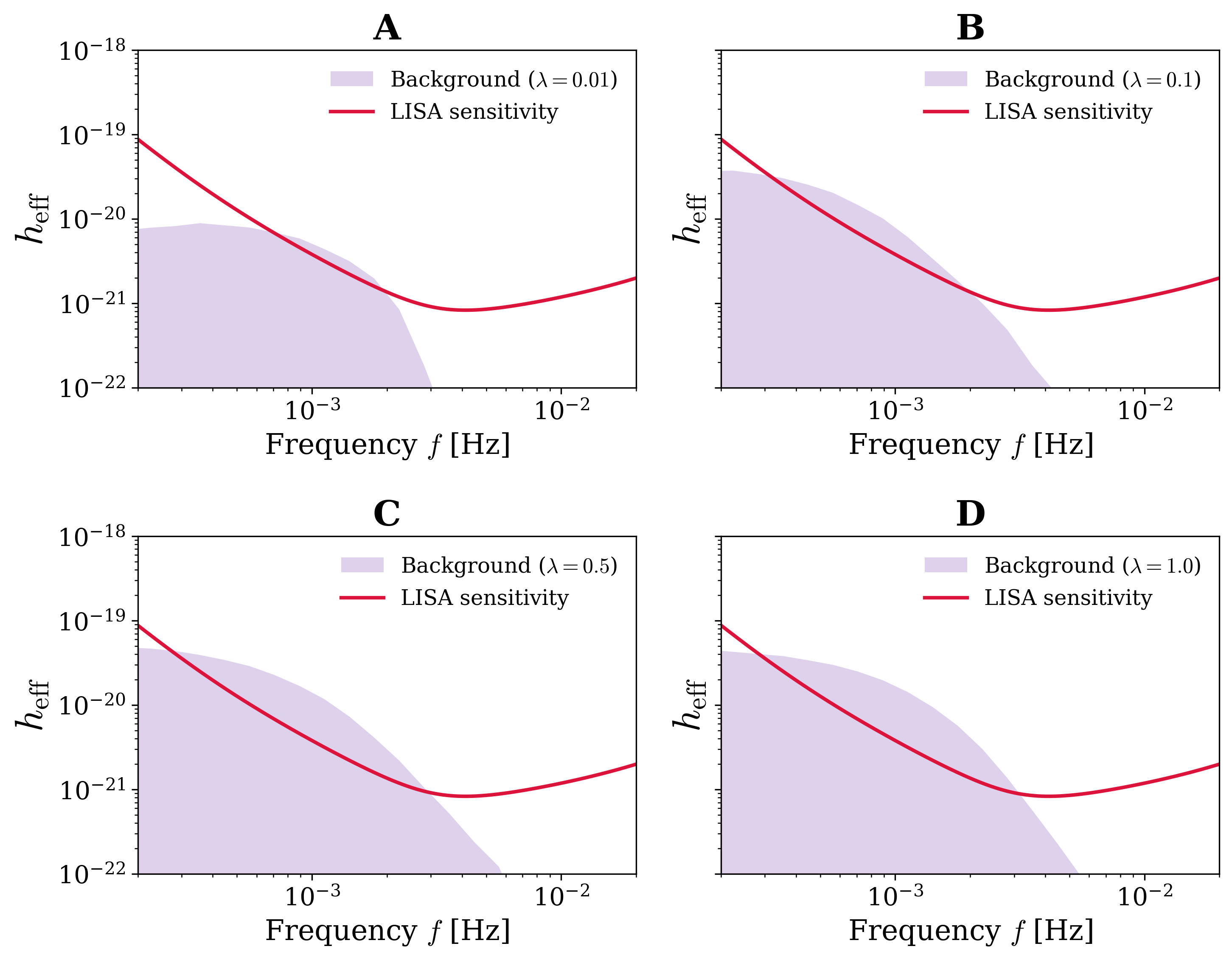}
    \caption{GW strain vs. frequency for models with varied CE prescriptions.}
    \label{fig1}
\end{figure*}

Model B of Figure \ref{fig1} shows the effective characteristic strain spectrum, $h_\mathrm{eff}(f)$, of the unresolved GW background from this population, scaled to the present Galactic population of $10^8$ BWDs. The spectrum exhibits the familiar Galactic background shape that shows a rising power law at low frequencies, flattening near the turnover around $1$\,mHz, and declining towards higher frequencies as the number of systems decreases. In this model, the signal amplitude at the knee is around $10^{-20}$, lying well above the \textit{LISA} instrumental noise. 

Most of the power in the background comes from binary systems with orbital periods of minutes to a few hours, which evolve slowly under GW emission and contribute nearly monochromatic signals during \textit{LISA}'s lifetime. The more compact systems contribute at higher frequencies but are fewer in number, leading to the rapid decline of $h_\mathrm{eff}(f)$ above this range. This fiducial spectrum serves as our baseline for comparison with other evolutionary models in the following sections, where we vary key binary evolution parameters to study their impact on the shape and amplitude of the GW background. We summarise all these models in Table~\ref{tab1}.

\begin{table*}
\centering
\caption{Summary of the binary evolution models considered in this work.}
\label{tab1}
\begin{tabular}{lccccc}
\hline
Model & $\alpha_{\rm CE}$ & $\lambda_{\rm CE}$ & $\beta$ & $Z$ & AM loss \\
\hline
A            & 1 & 0.01 & Thermal & Solar & Isotropic \\
B (Baseline) & 1 & 0.1  & Thermal & Solar & Isotropic \\
C            & 1 & 0.5  & Thermal & Solar & Isotropic \\
D            & 1 & 1.0  & Thermal & Solar & Isotropic \\
E            & 1 & 0.1  & 0.1     & Solar & Isotropic \\
F            & 1 & 0.1  & 0.5     & Solar & Isotropic \\
G            & 1 & 0.1  & Thermal & 0.0002 & Isotropic \\
H            & 1 & 0.1  & Thermal & 0.002 & Isotropic \\
I            & 1 & 0.1  & Thermal & Solar & Jeans \\
J            & 1 & 0.1  & Thermal & Solar & Arbitrary(=2) \\
\hline
\end{tabular}
\end{table*}

\subsection{Models with varied CE prescriptions}
To assess the impact of CE physics on the predicted Galactic background, we hold the CE efficiency fixed at $\alpha_{\rm CE} = 1.0$ and vary the envelope binding parameter $\lambda_{\rm CE}$. We consider four representative models: $\lambda_{\rm CE} = 0.01$ (Model A), $\lambda_{\rm CE} = 0.1$ (Model B; fiducial), $\lambda_{\rm CE} = 0.5$ (Model C), and $\lambda_{\rm CE} = 1.0$ (Model D). This sequence illustrates how assumptions about the envelope’s binding energy influence the survival of progenitors through CE evolution, and in turn, the frequency distribution of BWDs observable by \textit{LISA}.

Figure \ref{fig1} shows $h_{\rm eff}(f)$ for these four models. The differences in spectral shape can be directly traced to how the CE prescription alters the post-CE orbital separations.

\begin{itemize}
    \item Model A ($\lambda_\mathrm{CE}= 0.01$): A very small $\lambda$ corresponds to a tightly bound envelope, requiring more orbital energy to unbind. Most binaries fail to eject the envelope and merge prematurely, leaving very few surviving BWDs. The background spectrum is therefore strongly suppressed at all frequencies, and the contribution to the \textit{LISA} band is therefore strongly reduced.
    \item Model B ($\lambda_\mathrm{CE} = 0.1$, fiducial): This intermediate binding energy allows a reasonable fraction of binaries to survive. The resulting strain spectrum exhibits the canonical background shape: a steep rise at low frequencies where many wide binaries accumulate, a flattening near the turnover around $1$\,mHz, and a rapid decline above this point where only the tightest systems remain.
    \item Model C ($\lambda_\mathrm{CE} = 0.5$): With a less tightly bound envelope, less orbital energy is required for envelope ejection, so surviving systems generally emerge with wider post-CE separations. This shifts more power to the low-frequency end of the spectrum ($f \lesssim 1$\,mHz), broadening the background hump. At the turnover, the amplitude is slightly higher than in the fiducial model.
    \item Model D ($\lambda_\mathrm{CE} = 1.0$): Further increasing $\lambda$ again enhances binary survival, but the effect saturates. The spectrum closely resembles Model C, with nearly identical slope and high frequency behaviour. The main distinction is that Model D yields a marginally higher amplitude near the knee, reflecting the small additional contribution of survivors at shorter orbital periods. 
\end{itemize}

\begin{figure*}[htpb]
    \centering
    \includegraphics[width=\linewidth]{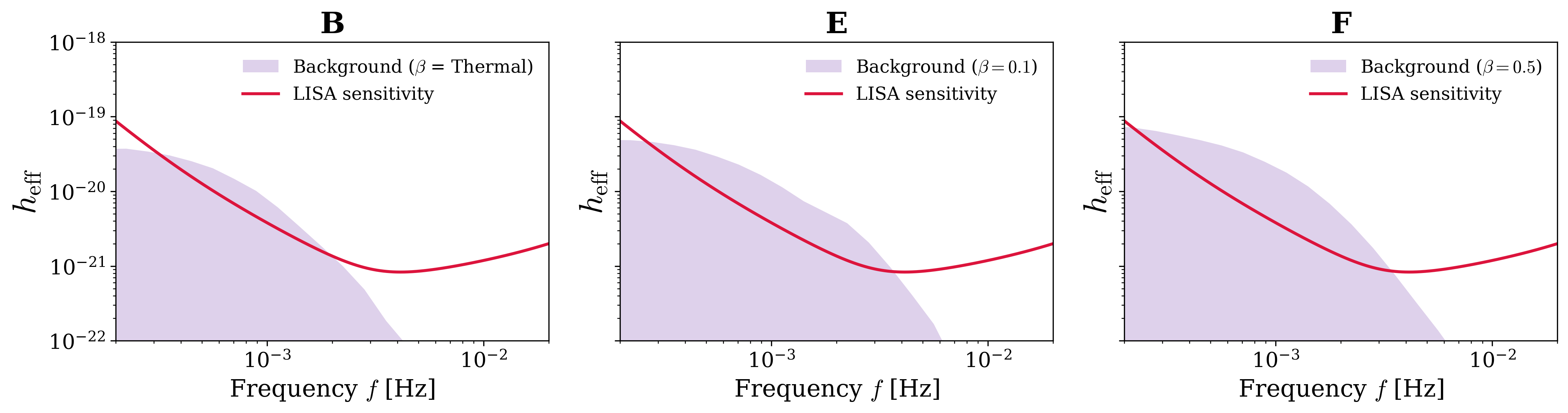}
    \caption{GW strain vs. frequency for models with varied MT prescriptions.}
    \label{fig2}
\end{figure*}

\subsection{Models with varied mass-transfer prescriptions}
Beyond the CE phase, the efficiency of stable MT is another key factor shaping the Galactic GW background. The outcome of RLOF depends on the fraction of transferred material retained by the accretor, quantified by the MT efficiency $\beta$. In this work, all non-conservative MT models adopt the isotropic re-emission prescription, in which the non-accreted material is expelled from the vicinity of the accretor and carries away the accretor's specific orbital AM. The combined effect of mass retention and AM loss determines whether the binary orbit contracts, expands, becomes unstable, or survives as a detached BWD. This orbital response directly imprints on the frequency distribution of BWDs observable in the \textit{LISA} band.

We explore three prescriptions: thermal-timescale MT (fiducial, Model B), a highly non-conservative case with $\beta=0.1$ (Model E), and an intermediate case with $\beta=0.5$ (Model F) with their corresponding $h_{\rm eff}(f)$ are shown in Figure \ref{fig2}. We also test a fully conservative case with \(\beta=1.0\). Its unresolved background lies below the \textit{LISA} instrumental sensitivity across the frequency range, hence we do not report it in further studies for clarity. 

\begin{itemize}
     \item Model B ($\beta$=thermal-timescale): In the fiducial model, the accreted fraction is set by the thermal response of the accretor. Excess material is expelled from the vicinity of the accretor, carrying away angular momentum. This produces moderate orbital shrinkage: enough to form a low frequency BWD background, but not enough to populate the most compact \textit{LISA} band systems efficiently. Therefore, the background is strongest at low frequencies and drops rapidly toward higher frequencies.
    \item Model E ($\beta=0.1$): With only $10\%$ of the mass accreted, most of the transferred material escapes the system. Under the adopted AM loss prescription, this strong mass loss tends to reduce orbital separations. While this channel does form compact binaries, it also increases the probability of premature mergers during mass transfer. As a result, the population that survives into the \textit{LISA} band is limited, and the background strain is weaker and more confined in frequency compared to the intermediate case of $\beta=0.5$.
    \item Model F ($\beta=0.5$): At $50\%$ accretion efficiency, the competing effects of orbital contraction and expansion are comparatively balanced. Enough mass is retained to avoid excessive AM loss, while sufficient material is still expelled to shrink orbits effectively. In our simulations, this prescription produces the largest effective population of BWDs in the \textit{LISA} frequency range, leading to the broadest spectrum and the highest effective strain. The spectral width in this case reflects the diverse range of post MT separations: wide binaries pile up at lower frequencies, while a significant number of compact systems extend power into the higher frequency regime.
\end{itemize}

\begin{figure*}[htpb]
    \centering
    \includegraphics[width=\linewidth]{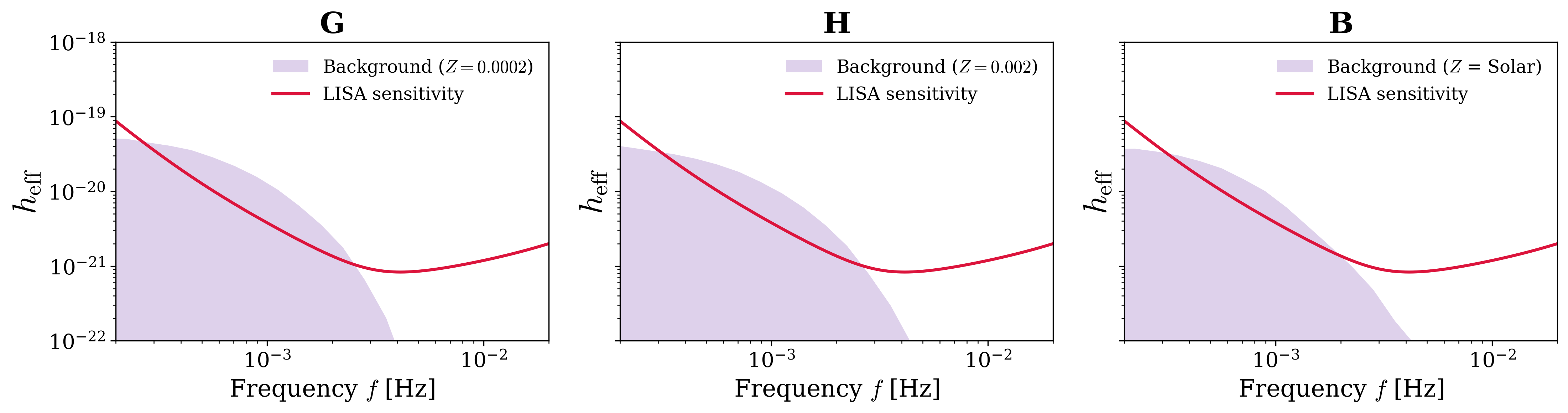}
    \caption{GW strain vs. frequency for models with varied metallicity prescriptions.}
    \label{fig3}
\end{figure*}

\subsection{Models with varied Metallicity prescriptions}
Metallicity is a key parameter in stellar evolution, shaping both the structure of stars and the outcome of binary interactions. Its influence arises mainly through two channels: (i) the strength of stellar winds, which scale with metallicity, and (ii) the radius of stars at different evolutionary stages. Metallicity affects wind mass loss, stellar radii, evolutionary timescales, and the onset of binary interactions. These effects propagate through binary evolution and change both the number of BWDs formed and their final orbital-period distribution.

We compare three metallicities: $Z=0.0002$ (Model G), $Z=0.002$ (Model H), and $Z=0.014$ (Model B; solar, fiducial). The resulting $h_\mathrm{eff}(f)$, shown in Figure \ref{fig3}, exhibit similar broad shape in all cases, a rise at low frequencies, flattening around $\sim 1$\,mHz, and decline at higher frequencies, but their normalization and location relative to the \textit{LISA} band differ systematically with metallicity. 

\begin{itemize}
    \item Model G ($Z=0.0002$): At extremely low metallicity, stellar winds are weak, so progenitors retain more mass before entering binary interaction phases. This favours the formation of a larger effective population of BWDs contributing in the \textit{LISA} band. In our simulations, this model produces the highest background amplitude among the metallicity models, indicating that a large number of compact BWDs are present across the relevant frequency range.
    \item Model H ($Z=0.002$): The intermediate metallicity model also benefits from relatively weak winds, but less strongly than the extremely low metallicity case. As a result, it produces a broad background spectrum with an amplitude below that of Model G. This suggests a smaller effective population of compact BWDs radiating in the millihertz band, while still maintaining a substantial \textit{LISA} band contribution.
    \item Model B ($Z=0.014$, solar): At solar metallicity, stronger stellar winds remove more mass from the progenitors and modify the subsequent binary interactions. Under the adopted binary evolution prescriptions, this leads to fewer effective compact BWDs contributing to the \textit{LISA} band. Consequently, the background amplitude is suppressed relative to the low metallicity models, although the overall spectral shape is preserved.
\end{itemize}

\begin{figure*}[htpb]
    \centering
    \includegraphics[width=\linewidth]{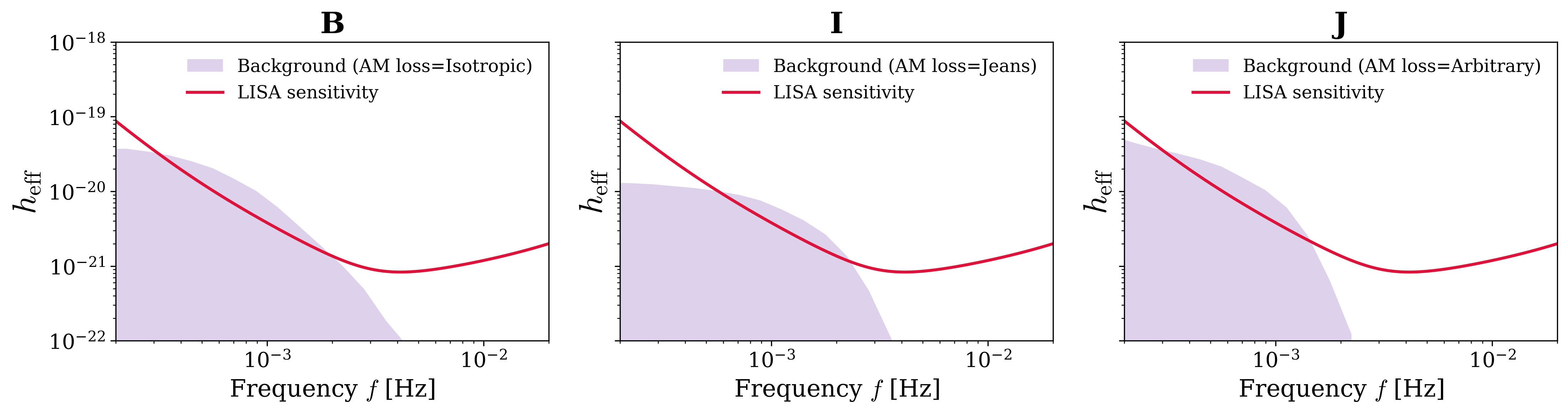}
    \caption{GW strain vs. frequency for models with varied AM loss prescriptions.}
    \label{fig4}
\end{figure*}

\subsection{Models with varied Angular Momentum loss prescriptions}
To investigate how different assumptions about AM loss during non-conservative MT influence the GW background, we fix the CE parameters at $(\alpha_{\rm CE}, \lambda_{\rm CE}) = (1.0, 0.1)$, MT parameter at $\beta=\text{Thermal}$ and vary the AM loss prescription. We examine three representative prescriptions that correspond to distinct limiting cases for the origin of the escaping material: loss from the vicinity of the accretor, loss directly from the donor, and a parameterized enhanced AM-loss channel not tied to either star. In the isotropic re-emission case (Model B), mass is transferred from the donor to the accretor, but only a part of it is retained. The remaining material is expelled from the vicinity of the accretor and therefore carries away the specific orbital AM of the accretor. In the Jeans mode case (Model I), mass is not first transferred to the companion; instead, it is lost directly from the donor as an isotropic wind. The escaping material therefore carries the specific orbital AM of the donor. In the arbitrary AM loss case (Model J), material is lost from the binary system and is not tied to either star. Here, each unit of lost mass removes twice the average specific orbital AM of the binary. This model provides a parameterized enhanced AM loss case, which can approximate uncertain channels such as circumbinary outflows or loss through regions not tied to either star. It also allows us to test whether the predicted BWD population and GW background are sensitive to stronger AM removal than in the donor or accretor centered prescriptions. These prescriptions control how much orbital AM is removed during stable MT, thereby affecting the orbital separations, merger times, and lifetimes of the resulting BWDs.

Figure \ref{fig4} presents the corresponding effective strain spectra, $h_{\mathrm{eff}}(f)$, for the three AM loss models, alongside the \textit{LISA} sensitivity curve. While all three curves share the characteristic structure of the unresolved BWD background, modest systematic differences appear.

\begin{itemize}
    \item {Model B (Isotropic re-emission):} In this case, the non-accreted material is expelled from the vicinity of the accretor and carries the accretor's specific orbital AM. This generally leads to a moderate removal of orbital AM, allowing many systems to survive stable MT and form BWDs. The resulting GW background shows the typical confusion spectrum, with strong power at low frequencies, a broad maximum around the mHz range, and a decline at higher frequencies where only the most compact systems contribute.
    \item {Model I (Jeans mode):} Here, mass is lost directly from the donor as an isotropic wind and carries the donor's specific orbital AM. Compared with isotropic re-emission, this prescription usually removes less orbital AM during MT, leading to wider post MT binaries. As a result, fewer systems evolve into close, high frequency BWDs, shifting the background power toward lower frequencies and reducing the strain amplitude around the central \textit{LISA} band.
    \item {Model J (Arbitrary AM loss, $\gamma=2$):} In this model, the lost material is assigned a fixed specific AM equal to twice the average specific orbital AM of the binary. This represents a stronger and more parameterized AM loss prescription, not directly tied to either the donor or the accretor. It can therefore produce a different post MT separation distribution, affecting both the overall background amplitude and the position of the high frequency cut off.
\end{itemize}

\subsection{Statistical comparison for the models}
To quantify the differences between the fiducial model and the alternative binary evolution prescriptions, we compute a reduced $\chi^2$ statistic using the realization to realization scatter as the uncertainty. This scatter based statistic provides a conservative measure of distinguishability, since it accounts for stochastic variations between different Galactic realizations. The comparison is performed over the same logarithmically rebinned frequency range used for the PSD analysis.

\begin{table}
\centering
\caption{$\chi_{\rm red}^2$ comparison between the baseline model and alternative binary evolution prescriptions, assuming a 4\,yr observing time and 100 Galactic realizations.}
\label{tab2}
\begin{tabular}{lccccc}
\hline
Model set & Variant & $\chi^2_{\rm red}$ \\
\hline
CE 
& $\lambda_{\rm CE} = 0.01$ & 7.78 \\
& $\lambda_{\rm CE} = 0.5$  & 2.63 \\
& $\lambda_{\rm CE} = 1.0$  & 4.27 \\
\hline
MT accretion efficiency 
& $\beta = 0.1$ & 5.31 \\
& $\beta = 0.5$ & 7.64 \\
\hline
Metallicity 
& $Z = 0.0002$ & 2.91 \\
& $Z = 0.002$  & 1.12 \\
\hline
AM loss 
& Jeans mode & 3.09 \\
& Arbitrary, $\gamma = 2$ & 0.10 \\
\hline
\end{tabular}
\end{table}

The resulting $\chi_{\rm red}^2$ values are summarized in Table~\ref{tab2}. The CE models all yield $\chi^2_{\rm red}>1$, indicating measurable deviations from the fiducial $\lambda_{\rm CE}=0.1$ model. Among these, the $\lambda_{\rm CE}=0.5$ model is closest to the baseline, while the $\lambda_{\rm CE}=1.0$ and $\lambda_{\rm CE}=0.01$ models show stronger deviations. This confirms that the unresolved Galactic BWD background is sensitive to the assumed CE prescription.

The MT efficiency models also produce large deviations, especially $\beta=0.5$, reflecting the strong enhancement and broadening of the background seen in the corresponding spectrum. The $\beta=0.1$ model also differs substantially from the fiducial case, demonstrating that the efficiency of stable MT has a strong effect on the number and period distribution of compact BWDs in the \textit{LISA} band.

The metallicity models show a more moderate level of distinguishability. The very low metallicity model, $Z=0.0002$, differs appreciably from the solar metallicity fiducial model, whereas the $Z=0.002$ model remains close to the baseline. This indicates that the BWD background is sensitive to metallicity, but that moderate metallicity changes may be partially degenerate with realization to realization scatter and other population assumptions.

The AM loss prescriptions produce comparitively weaker deviations. The Jeans mode gives a good difference relative to the isotropic baseline, while the arbitrary $\gamma=2$ model has $\chi^2_{\rm red}<1$. A value below unity indicates that the difference between the two spectra is smaller than the adopted realization scatter over the frequency range considered.

\begin{figure*}[htpb]
    \centering
    \includegraphics[width=\linewidth]{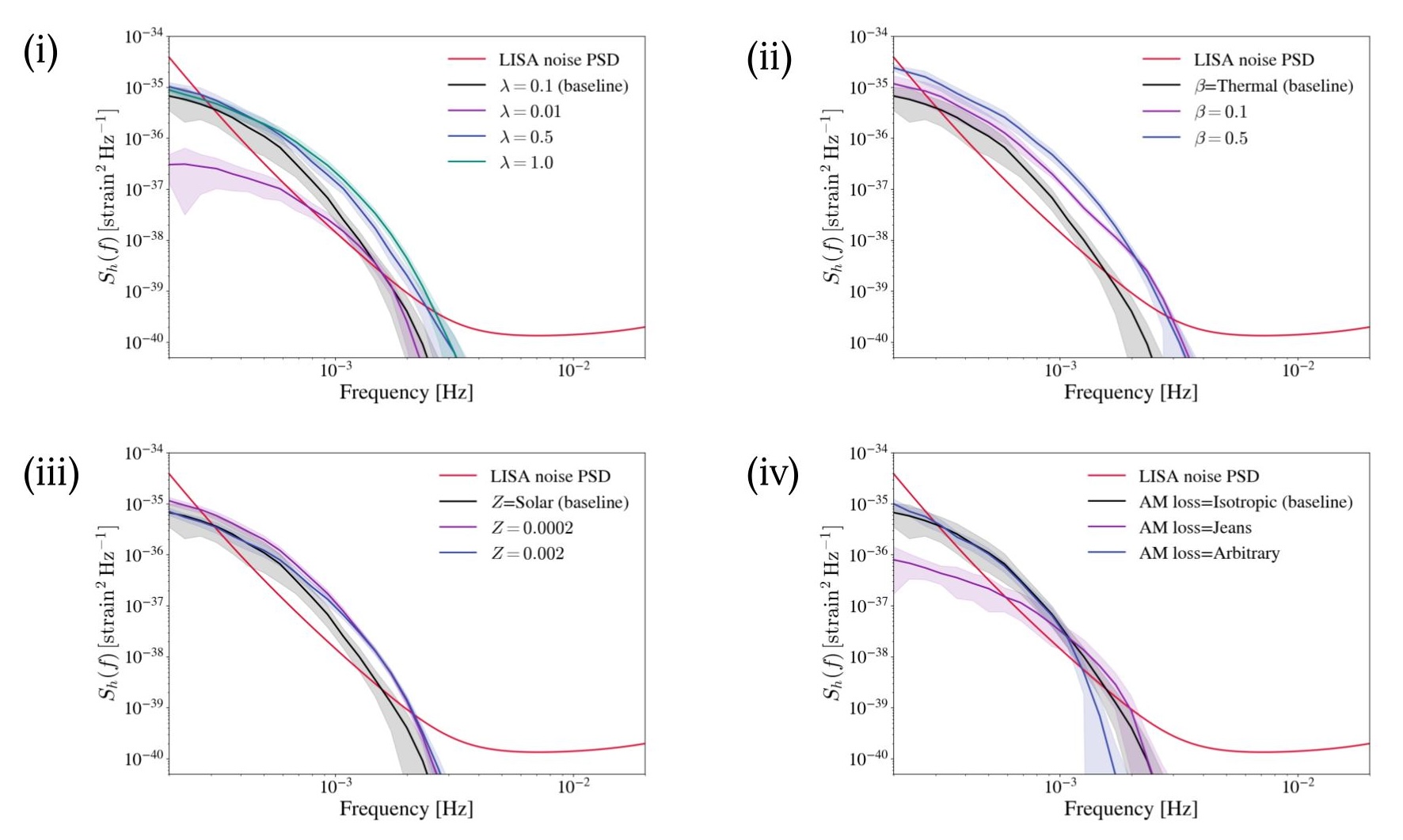}
    \caption{Figure (i), (ii), (iii), and (iv) show the PSDs for CE, MT, Metallicity, and AM loss models, respectively.}
    \label{Fig. chi-square}
\end{figure*}

Overall, the $\chi^2$ comparison shows that the \textit{LISA} BWD background is not equally sensitive to all uncertain binary evolution parameters, evident in Figure \ref{Fig. chi-square}. In our models, the strongest imprints arise from stable MT efficiency and CE evolution, followed by metallicity, while AM loss prescriptions leave comparatively weaker or more degenerate signatures. This hierarchy is important for interpreting future \textit{LISA} measurements, as it identifies which aspects of binary evolution are most likely to be constrained by the unresolved Galactic background.

\section{Discussion}{\label{Sec7}}

\subsection{Comparison with previous works}
The Galactic GW background produced by BWDs has been studied extensively since the early works of \cite{1987ApJ...323..129E,1987A&A...176L...1L,1990ApJ...360...75H}. These works, based on simplified assumptions for binary formation and Galactic structure, predicted relatively strong background signals. Subsequent population synthesis models have progressively refined these estimates using more realistic treatments of binary evolution.

A key milestone was the self-consistent population synthesis study of \cite{2001A&A...375..890N} assuming a constant SFR over a Galactic disk age of $13-15$\,Gyr. Their fiducial model with $\alpha_\mathrm{CE}\simeq1$, yielded a background peaking at $f\simeq1.6-2$\,mHz and $h_c\simeq10^{-20}$, about a factor of three lower than earlier analytic predictions, and confirmed that semi-detached systems contribute negligibly. More recent studies \citep{2022MNRAS.511.5936K,2010A&A...521A..85Y}, spanning $\alpha_\mathrm{CE}=0.2-1.0$ and $Z=0.0002-0.02$, consistently find peaks at around $2-4$\,mHz and amplitudes $h_c \sim 10^{-20}-10^{-19}$.

Our fiducial model (model B) produces a background peaking near $\sim$3\,mHz with $h_\mathrm{eff}\approx 10^{-20}$. This lies within 0.2\,dex of the results reported by \cite{2001A&A...375..890N,2022MNRAS.511.5936K}, confirming our assumptions yield a realistic and physically consistent representation of the dominant Galactic GW background expected in the \textit{LISA} frequency band.

\subsection{Physical interpretation of the model variations}

Our results show that the unresolved Galactic BWD background is sensitive to several aspects of binary evolution and different parameters affect the spectrum in physically distinct ways. The CE prescription primarily controls the survival probability of interacting progenitors and the post-CE orbital separations of the systems that avoid merger. In our CE model sequences, the smallest $\lambda_{\rm CE}=0.01$ suppresses the background strongly by making envelope ejection difficult and increasing the fraction of systems that merge during the CE phase. By contrast, larger $\lambda_{\rm CE}$ values allow more binaries to survive and preferentially populate wider post-CE orbits, enhancing the low-frequency background. The convergence of the $\lambda_{\rm CE}=0.5$ and $\lambda_{\rm CE}=1.0$ cases suggests that once the envelope becomes sufficiently easy to unbind, the background response begins to saturate.

The stable MT prescription affects the background through a different mechanism. Here the key competition is between orbital contraction caused by non-conservative mass and AM loss, and orbital widening caused by conservative mass exchange. Our results indicate that highly non-conservative MT suppresses the \textit{LISA} band population by promoting mergers or overly aggressive orbital shrinkage. The fully conservative case, although not included in the quantitative comparison, acts as a limiting example in which orbital widening strongly suppresses the compact BWD population in the \textit{LISA} band. Intermediate transfer efficiencies produce the broadest and strongest background because they balance these competing effects and generate a wide range of BWD separations.

Metallicity regulates the efficiency with which progenitor binaries produce compact BWDs through its influence on stellar winds, stellar radii, evolutionary timescales, and the timing of binary interactions. We find that lower metallicity leads to a larger effective number of compact BWDs contributing in the \textit{LISA} band, while solar metallicity produces a lower background amplitude. This appears primarily as a change in the overall background normalization, although the spectral shape remains broadly similar.

The AM loss prescription during non-conservative MT produces a more subtle effect. Changing the specific AM carried away by expelled material modifies the post MT separation distribution, but the resulting spectra remain broadly similar in overall morphology. The Jeans mode favours wider post MT systems and produces a weaker, softer spectrum relative to the isotropic baseline. The arbitrary enhanced loss model changes the spectral width and high frequency cut off, but its overall background remains close to the fiducial morphology. This suggests that AM loss assumptions may be more difficult to isolate from the unresolved background alone than CE evolution or MT efficiency.

\subsection{Interpretation of the statistical analysis}

The $\chi^2$ comparison provides a quantitative measure of how strongly the PSD predicted by different models departs from the fiducial case. We use the realization to realization scatter as the variance estimate, so the resulting $\chi_\mathrm{red}^2$ provide a conservative measure of model distinguishability in the presence of stochastic fluctuations between Galactic realizations.

For the CE model sequence, all alternative prescriptions yield $\chi_\mathrm{red}^2>1$. The $\lambda_{\rm CE}=0.5$ model is the closest to the fiducial $\lambda_{\rm CE}=0.1$ case, while the $\lambda_{\rm CE}=0.01$ and $\lambda_{\rm CE}=1.0$ models are more clearly separated. This ordering mirrors the spectral behaviour seen in the background curves: moderate changes in the CE binding energy prescription perturb the background less strongly than more extreme assumptions. The CE results therefore indicate that the unresolved Galactic BWD background is sensitive to the assumed treatment of envelope binding energy.

Similarly, larger $\chi_\mathrm{red}^2$ are obtained for the MT efficiency models too, especially the $\beta=0.5$ case. This is consistent with the strong enhancement and broadening of the background spectrum in this model. The highly non-conservative $\beta=0.1$ case also differs substantially from the fiducial model, showing that the stable MT prescription can strongly affect the number and orbital-period distribution of compact BWDs in the \textit{LISA} band.

The metallicity models show a more moderate level of separation. The very low metallicity model, $Z=0.0002$, produces a measurable departure from the solar metallicity fiducial case, whereas the $Z=0.002$ model remains much closer to the baseline. This suggests that the background is sensitive to metallicity, but that moderate metallicity variations may be partially degenerate with realization scatter and other population assumptions.

By contrast, the AM loss prescriptions produce weaker deviations. The Jeans mode gives a modest departure from the isotropic baseline, while the arbitrary enhanced loss model remains close to the fiducial spectrum. In particular, $\chi_\mathrm{red}^2<1$ indicates that the difference from the baseline is smaller than the adopted realization to realization scatter over the frequency range considered.

More generally, the $\chi^2$ analysis complements the visual comparison of spectra. While differences in background amplitude, slope, or turnover can often be seen by eye, $\chi_\mathrm{red}^2$ provides a statistical way to assess whether these differences are large compared with stochastic Galactic variance. It should therefore be interpreted as a first diagnostic of the discriminating power of the unresolved background, rather than as a full \textit{LISA} parameter inference forecast. Within this framework, our results suggest that the background is most sensitive to stable MT efficiency and CE evolution, moderately sensitive to metallicity, and less sensitive to the AM loss prescriptions considered here.

\subsection{Astrophysical implications for \textit{LISA}}
A central implication of this work is that the unresolved Galactic BWD background should not be viewed solely as a confusion limited signal for other \textit{LISA} sources. It is also an astrophysical signal in its own right. Because it arises from the superposition of a large and diverse population of binaries, it encodes integrated information about binary formation, interaction, and survival across the history of the MW.

This is particularly valuable because the unresolved background probes a part of the BWD population that cannot be fully accessed through individually resolved systems alone. Resolved sources will provide direct information on the loudest and typically most compact binaries, but the background contains contributions from the much larger ensemble of weaker and more numerous systems. The two observables are therefore complementary: resolved detections constrain individual source properties, while the background constrains the collective population and its evolutionary pathways.

Our results suggest that background measurements may be especially informative about processes that control binary survival and orbital shrinkage, such as CE ejection and AM loss during MT. The sensitivity of the background to metallicity further raises the possibility that Galactic stellar-population properties may also be encoded in the spectrum, although likely with stronger degeneracies. In the longer term, combining the unresolved background with resolved BWD catalogues and electromagnetic constraints on the Galactic stellar population could provide a powerful route to constraining binary evolution physics in the MW.

\subsection{Model limitations and caveats}
Several simplifications in our analysis should be considered when interpreting the results. First, the emitted GW signal is approximated using the dominant quadrupole harmonic of a circular binary. This is a reasonable approximation for most Galactic BWDs, which are expected to have negligible eccentricities by the time they enter the \textit{LISA} band.

Second, our background construction uses simplified prescriptions for detectability and the power spectral density (PSD). Resolved and unresolved binaries are separated using a monochromatic SNR threshold based on an analytic \textit{LISA} sensitivity curve. Similarly, each source is treated as a stationary monochromatic sinusoid in the PSD calculation. We therefore do not consider here the full time dependent \textit{LISA} response, sky modulation, frequency evolution, and a realistic iterative source subtraction procedure. These approximations are sufficient for comparing relative trends across binary evolution models, although they do not constitute a full end to end forecast of \textit{LISA} data analysis performance.

Third, the Galactic population model is simplified. We adopt an effective MW normalization, approximate a constant star formation history using discrete bursts, and assume a fixed spatial distribution. The MW has a more complex assembly history, metallicity evolution, and spatially varying stellar populations across the thin disk, thick disk, bulge, and halo. These features may affect both the amplitude and anisotropy of the background and may introduce additional degeneracies with binary evolution parameters. Incorporating a more detailed Galactic model is beyond the scope of the present study and will be addressed in future work.

Finally, the binary evolution prescriptions explored here are designed to isolate the effects of key physical assumptions. Within this controlled framework, our results highlight how the unresolved background responds to changes in the underlying binary evolution physics.

To summarise, our results demonstrate the unresolved Galactic BWD background is an integrated population diagnostic with different evolutionary prescriptions leave different signatures in the normalization, width, slope, and turnover of the spectrum. This makes the background a potentially useful probe of binary evolution physics.

\begin{acknowledgements}
    S.K., T.B., and D.R. acknowledge funding by the National Center for Science, Poland, grant no. 2023/49/B/ST9/02777.
\end{acknowledgements}

\section*{Data availability statement}
The data used in this article are obtained from \textsc{compas} simulation, which will be shared on reasonable request to the corresponding author.

\bibliographystyle{aa}
\bibliography{Bibliography}

\end{document}